# Analysis of Thermoelectric Properties of Scaled Silicon Nanowires Using an Atomistic Tight-Binding Model

Neophytos Neophytou[1], Martin Wagner[2], Hans Kosina[1], and Siegfried Selberherr[1]

[1]Institute for Microelectronics, Technische Universität Wien, Vienna, Austria

[2]O-Flexx Technologies GmbH, Ratingen, Germany

## Abstract

Low dimensional materials provide the possibility of improved thermoelectric performance due to the additional length scale degree of freedom for engineering their electronic and thermal properties. As a result of suppressed phonon conduction, large improvements on the thermoelectric figure of merit, $ZT$, have been recently reported in nanostructures, compared to the raw materials' $ZT$ values. In addition, low dimensionality can improve a device's power factor, offering an additional enhancement in $ZT$. In this work the atomistic $sp^3d^5s^*$-spin-orbit-coupled tight-binding model is used to calculate the electronic structure of silicon nanowires (NWs). The Landauer formalism is applied to calculate an upper limit for the electrical conductivity, the Seebeck coefficient, and the power factor. We examine n-type and p-type nanowires of diameters from 3nm to 12nm, in [100], [110], and [111] transport orientations at different doping concentrations. Using experimental values for the lattice thermal conductivity in nanowires, an upper limit for $ZT$ is computed. We find that at room temperature, scaling the diameter below 7nm can at most double the power factor and enhance $ZT$. In some cases, however, scaling does not enhance the performance at all. Orientations, geometries, and subband engineering techniques for optimized designs are discussed.

**Index terms:** thermoelectric, conductivity, tight-binding, atomistic, $sp^3d^5s^*$, Landauer, Seebeck coefficient, silicon, nanowire, $ZT$.

# I. Introduction

The progress in nanomaterials' synthesis has allowed the realization of low-dimensional thermoelectric devices based on materials such as one-dimensional nanowires, thin films, and two-dimensional superlattices [1, 2, 3, 4]. The ability of a material to convert heat into electricity is measured by the dimensionless figure of merit $ZT=\sigma S^2 T/(k_e+k_l)$, where $\sigma$ is the electrical conductivity, $S$ is the Seebeck coefficient, and $k_e$ and $k_l$ are the electronic and lattice part of the thermal conductivity, respectively. Low-dimensional materials offer the capability of improved thermoelectric performance due to the additional length scale degree of freedom on engineering $S$, $\sigma$, and $k_l$. The introduction of the length scale as a design parameter can provide partial control over the dispersions and scattering mechanisms of both electrons and phonons.

Recently, large improvements in $ZT$ have been reported in silicon nanowires ($ZT$ approx. 1) compared to the bulk $ZT$ value of silicon ($ZT_{bulk}$ approx. 0.01) [1, 2]. Most of this improvement has been a result of suppressed phonon conduction ($k_l$). As suggested by various studies, however, low dimensionality can be beneficial for increasing the power factor ($\sigma S^2$) of the device as well [5, 6]. The sharp features in the low-dimensional density of states (*DOS*) as a function of energy can improve $S$ [5, 6], as this quantity is proportional to the energy derivative of the *DOS*. In addition, it was theoretically shown that structural quantization can potentially improve $\sigma$ through reduction of the material's effective mass [7, 8]. A proper choice of the transport orientation can also improve $\sigma$. Of course an improvement in $\sigma$ can degrade $S$, since these two quantities are inversely related. At the nanoscale, however, subband engineering techniques can be used to optimize this interrelation and maximize the power factor. The enhancement in $ZT$ above a given material's bulk value, which was reported in [1] and [2] can, therefore, be increased even further by optimizing the numerator of $ZT$.

In this work the atomistic $sp^3d^5s^*$-spin-orbit-coupled tight-binding model [9, 10, 11] is used to calculate the electronic structure of cylindrical silicon nanowires. The Landauer formalism [12] is then applied to calculate an upper limit for the electrical conductivity, the Seebeck coefficient, and the power factor as proposed by Kim *et al.* [13]. The calculations in this work are performed with realistic electronic structures for NWs and the effects of orientation and diameter are thoroughly examined. The Landauer formalism employed, is an ideal, coherent electronic transport model, and excludes scattering mechanisms and non-idealities. In that way, it is different from the traditional diffusive, incoherent electrical Boltzmann transport models. Therefore, the results presented here, should not be interpreted as providing absolute, realistic device performance numbers, but as exploring the upper limit of the performance and at which length scales band engineering can improve the power factor. Our objectives are: i) to examine, if in principle diameter scaling can enhance the power factor in nanostructures using realistic dispersions, and if so, ii) at which length scale improvements can be expected, iii) what is the theoretical upper limit that can be achieved (can *ZT* larger than unity be achieved?), and iv) by investigating a large variety of NW categories identify the subband features that control those improvements. We examine n-type and p-type nanowires of diameters *D* from 3nm to 12nm, at different doping concentration levels. We also investigate [100], [110], and [111] transport orientations as shown in Fig. 1 (cross sections), since studies have suggested that the thermoelectric properties of NWs depend on orientation [14]. Finally, using experimental values for the lattice thermal conductivity in nanowires, an upper limit for the *ZT* values is computed.

We find that at room temperature, the power factor and *ZT* can be improved for NW diameters smaller than approx. 7nm. For larger diameters, however, these quantities saturate to lower values. The maximum benefits of scaling the diameter down to *D*=3nm, as calculated in this work, is to double the power factor. This, however, is not always the case. Reduced diameter size can offer no advantage, or even degrade the power factor in certain cases. Since the effect of surface roughness scattering tends to increase as the NWs' diameter decreases, dimensionality benefits on the power factor can be offset by scattering effects and thus, may, or may not be achieved in realistic structures. Reduced dimensionality, however, still largely improves thermoelectric performance by drastically reducing thermal conductivity.

## II. Approach

The NWs' bandstructure is calculated using the 20 orbital atomistic tight-binding spin-orbit-coupled model ($sp^3d^5s^*$-SO) [9]. In this model each atom in the NW is described by 20 orbitals (including spin-orbit-coupling). The advantage of using an atomistic model is that the nanowire is built on the actual zincblende lattice, and each atom is properly accounted in the calculation. It accurately captures the electronic structure and the respective carrier velocities, and inherently includes the effects of quantization and different orientations. The $sp^3d^5s^*$-SO model was extensively used in the calculation of the electronic properties of nanostructures with excellent agreement to experimental observations on various occasions [11]. Details of the model are provided in [7, 9]. We consider here infinitely long, cylindrical silicon NWs. No relaxation is assumed for the nanowire surfaces.

The dependence of the electronic structure on the diameter of the NWs is shown in the dispersions of Fig. 2 and Fig. 3. Figure 2 illustrates the dispersions of n-type NWs in the three different orientations [100], [110], and [111] (only half of the *k*-space is shown). The left column shows the dispersions of the *D*=3nm wires, and the right the dispersions of the *D*=12nm wires. All the dispersions are shifted to the same origin *E=0eV* for comparison purposes. Differences in the shapes of the dispersions between wires of different orientations and diameters, in the number of subbands, as well as the relative differences in their placement in energy can be observed. These differences will result in different electronic properties. Figure 3 shows the same features for p-type NWs, where a similar principal behavior can be observed. The origin of the different features in each orientation and diameter is explained in detail in [7] for the n-type case and in [8] for the p-type case. The important result here is that different electrical characteristics are expected in each wire case due to the different electronic structures. Eighty bands are shown in the larger diameter NW figures (right columns), the same as the number of bands used in the calculations. In the case of the large diameter NWs, the details of each individual subband, are not as crucial in providing understanding for the NW properties. Rather, the placement of the various valleys in energy and their degeneracy (especially in the n-type case), as well as the envelopes and the curvature of the subbands (especially in the p-type case) provide more understanding towards NW thermoelectric properties.

The Landauer formalism [12] is used to extract the electrical conductivity, the Seebeck coefficient, the power factor, and the electronic part of the thermal conductivity for each wire using its dispersion.

In the Landauer formalism the current is given by:

$$J = -\frac{q}{L}\sum_{k>0} v_k f_1 - \frac{q}{L}\sum_{k<0} v_k f_2 \qquad (1a)$$

$$= -\frac{q}{L}\sum_{k>0} v_k (f_1 - f_2), \qquad (1b)$$

where $v_k$ is the bandstructure velocity, and $f_1, f_2$ are the Fermi functions of the left and right contacts, respectively. An intermediate function $R^{(\alpha)}(f_1, f_2, T)$ can be defined as:

$$R^{(\alpha)} = \frac{-\frac{q}{L}\sum_{k>0} v_k (f_1 - f_2)(E_k - \mu_1)^\alpha}{(\mu_1 - \mu_2)}, \qquad (2)$$

where $\mu_1, \mu_2$ are the contact Fermi levels, and $E_k$ is the subband dispersion relation. This formula is the same as the one described in [13, 14], where for small driving fields $\Delta V$, $f_1 - f_2 = -q\Delta V \frac{\partial f_1}{\partial E}$ holds. Here, however, the computation is explicitly performed in $k$-space rather than energy-space. From this function, the conductance $G$, the Seebeck coefficient $S$, and the electronic part of the thermal conductivity $k_e$, can be derived as:

$$G = R^{(0)}, \qquad (3a)$$

$$S = \frac{1}{T}\frac{R^{(1)}}{R^{(0)}}, \qquad (3b)$$

$$k_e = \frac{1}{T}\left[R^{(2)} - \frac{[R^{(1)}]^2}{R^{(0)}}\right]. \qquad (3c)$$

The Landauer formalism as used here describes electronic transport in the ballistic, fully coherent limit. This is a different method from the Boltzmann transport method that describes diffusive, incoherent electrical transport. In principle, to include scattering the conductivity should be scaled by $\lambda(E)/L$, where $\lambda(E)$ is the energy dependent mean free path of carriers in each silicon nanowire, and $L$ is the length of the device as discussed by Kim [13]. An elementary treatment of scattering can therefore by incorporated within the Landauer formalism. The mean free path is a complicated function of scattering mechanisms, including the effect of surface roughness scattering, which is beyond the scope of this work. The effects of the carriers' finite mean free path and of the carriers' relaxation times as sketched in [14] and [15] are not treated here. Ballistic transport, however, cannot be achieved in a thermoelectric device. Our results indicate the *ideal* and *relative* change in performance as NWs of the same category are scaled from "bulk" like, large diameter NWs, to NWs of few nanometers in diameter. Our method uses information based only on the effects of bandstructure, and provides insight into design and optimization strategies for the bandstructure of thermoelectric devices. Especially, the Seebeck coefficient, which at first order is independent of scattering, depends most sensitively on the subband features of the dispersions.

## III. Results and Discussion

Figure 4 shows the extracted thermoelectric features for n-type [100] NWs, plotted as a function of the one-dimensional carrier concentration (corresponding to the doping concentration) as the Fermi level is pushed at higher energies into the subbands (see Fig. 2a and 2b). The NWs' diameters start from $D=3$nm (solid-black) to $D=12$nm (dot-black), and the blue lines indicate results for NWs with 1nm increment in diameter. The electrical conductivity (defined as $\sigma=G$/Area) of the smaller wires in Fig. 4a is shifted to the left compared to the conductivity of the larger wires. The reason is that at the same one-dimensional carrier concentration, the Fermi level is pushed at higher energies into the subbands of the narrower wires faster than in the case of the thicker NWs. In the case of the thicker wires, the Fermi level remains lower in energy (the larger number of subbands provides states for the required carrier concentration). The Seebeck coefficient on the other hand, in Fig. 4b, is higher (shifted to the right) for larger diameter NWs at the same one-dimensional carrier concentration than in the narrower NW cases. The reason is that there are many more subbands which are more spread in energy. The Seebeck coefficient is proportional to $E_F-E_C$, which increases as the subbands are spread in energy [13]. The power factor $\sigma S^2$, however, as shown in Fig. 4c, is favored for the smaller diameter NWs, for which the peak is almost twice as high as that of the thicker diameter ones. As the diameter increases from $D=3$nm to $D=7$nm the power factor reduces. For diameters larger than approx. 7nm, the peak of the power factor saturates, which indicates that in an ideal situation, performance benefits due to dimensionality will only be observed for NWs with diameters below 7nm. Similar effects are observed for the other NW family types we consider (n-type, p-type in different orientations). In reality, however, enhanced surface roughness scattering will reduce the conductivity more drastically in NWs with smaller diameters, and benefits may or may not be observed. We mention here that alternatively, one can plot $\sigma$, $S$ and $\sigma S^2$ as a function of the three-dimensional doping concentration (instead of one-dimensional). This will result in a shift of the relative positions of each curve on the *x*-axis depending on the NWs' area. The magnitude of the power factor peaks of wires with different diameters, however, does not change. In this case, the peaks appear less spread in the *x*-axis as when plotted against the one-dimensional doping values.

The same calculations are performed for differently orientated n-type and p-type NWs. The power factor, however, is determined by the interplay between $S$ and $\sigma$, and its peak appears at different carrier concentrations. In order to properly compare $\sigma$ and $S$ for the wires, we perform the comparison at the carrier concentration at which the peak of the power factor is observed. Figure 5a shows how the peak of the power factor varies for the n-type and p-type NWs with [100], [110], and [111] orientations as a function of the diameter. Figures 5b and 5c show the $\sigma$ and $S$ at the same carrier concentration at which the peak of the power factor appears. At larger diameters differences of the order of 30% can be observed for the different orientations. Much larger variations, however, are observed as the diameter is reduced below approx. 7nm, in which case the power factors for the n-type and the [100] p-type NWs increase substantially. In view of the fact that the maximum benefit in power factor observed under the ideal conditions considered here is 100%, the packing density of nanowire based thermoelectric devices must be at least 50%, as also discussed by Kim *et* al. [13], if these advantages on the power factor are to be utilized.

Recent reports on thermal conductivity measurements have shown that the thermal conductance of NWs scaled down to 15nm or 20nm in diameter can be reduced by two orders of magnitude from its bulk material value, and can be as low as $k$=1-2W/mK [1, 16, 17]. This reduction is responsible for the enhancement in the thermoelectric figure of merit ($ZT$) of silicon NWs, which was measured to be close to unity. Using the measured data $k_l$=2W/mK of the lattice thermal conductance of the $D$=15nm NW, we estimate the expected $ZT$ using the calculated upper limit of the power factor for the NWs considered in this study. The electronic part of the thermal conductivity, $k_e$, is used as calculated by Eq. 3c. The results for all nanowire families examined here are shown in Fig. 6. For clarity we only show $ZT$ for 3nm and 12nm wire diameters. In all cases $ZT$ values larger than unity can ideally be achieved. The smaller diameter NWs provide the larger $ZTs$, with large differences between the different orientations. As the diameter increases, the $ZT$ values decrease, and orientation differences become less important. The values are, however, above unity as well.

Since we consider ideal NWs and do not consider any scattering processes, surface roughness or structure relaxation which degrade the conductivity, our results provide the upper performance limit. The Seebeck coefficient will not be affected substantially, if scattering is included, but the conductivity will degrade. Surface roughness scattering, structure non-idealities, impurities, and phonon scattering are considered strong scattering mechanisms in NWs, and can cause large degradation of the conductance. Non-idealities, on the other hand, can increase $S$ by creating more singularities in the *DOS*. Consideration of scattering, however, is beyond the scope of this study. Kim *et* al. in Ref. [13], considered different scattering mechanisms, still within the Landauer framework for 1D, 2D and 3D devices, and concluded that individual conducting modes are used more efficiently in lower dimensions. Vo *et* al. in Ref. [14] have calculated $ZTs$ for 1.1 and 3nm diameter NWs considering scattering mechanisms and relaxation, and indeed they have found that $ZT$ values are lower from what we present for the $D$=3nm NWs, but can still remain above unity. This work, however, considers much larger wire diameters, and identifies the sizes at which dimensionality and orientation will play an effect. Our ballistic results for each NW category show the *relative* performance benefits as the diameter is scaled. For example, the performance of all n-type, and [100] p-type NWs improves as the diameter scales. On the other hand, in p-type [110] and [111] NWs (Fig. 6b), no benefits will be observed as the diameter is scaled. Actually, since scattering can be enhanced as the diameter scales, the performance can even be degraded.

Comparing the magnitude of the power factor and *ZT* for the n-type *D*=3nm NW case (Fig. 6a), the [100] NW, with a 4-fold degenerate Γ-valley (Fig. 2a), has a larger *ZT* than the [110] NW with a 2-fold degenerate Γ-valley. Subbands with higher degeneracies, or subbands with edges very close in energy, improve *S* and can be beneficial to the power factor. The n-type [111] NW has 6-fold degenerate valleys and the largest *ZT* in the ballistic case. The effective mass, *m\**, of this NW, however, is larger than that of the [100] and [110] NWs. The results of Vo *et* al. [14] show that if scattering is included, the *ZT* for the [111] wire will be degraded below the n-type [100] and [110] NWs. In that work, the increase in $\sigma$ and *S* due to the higher subband degeneracy, seems to be overcompensated by increased scattering due to the higher *m\**, which causes *ZT* to greatly decrease.

The results presented here are a consequence of the electronic structure properties alone. Bandstructure engineering can be beneficial to the power factor and design approaches can be identified. The approach followed needs to: i) Increase *S* by allowing more valleys nearby in energy, or using transport orientations with higher degeneracy subbands, as in the case of n-type [100] NWs which have higher performance than [110] NWs (Fig. 6a). ii) Keeping $\sigma$ high, by utilizing light effective mass subbands, or using strain engineering to reduce the effective masses of the subbands, however, not on the price of introducing subband splittings or reducing the degeneracies. Ideally the *DOS* should be increased through degeneracy increase, and not *m\** increase. In nanostructures, band engineering in this way is partially possible, especially when utilizing devices in different orientations, and benefits can therefore be achieved.

## IV. Conclusion

The upper limit of the thermoelectric power factor of silicon NWs was calculated for NWs of 3nm to 12nm in diameter using the Landauer formalism. The sp$^3$d$^5$s*-SO atomistic TB model was used for electronic structure calculation. n-type and p-type NWs in [100], [110], and [111] orientations were considered. Since scattering is not included in the calculations, our results show the *relative* change in performance due to bandstructure changes as the diameter of each NW category considered scales from *D*=12nm (bulk like) to *D*=3nm. It is shown that under ideal conditions, diameter scaling can enhance the power factor and *ZT* values of Si NWs by ~~even~~ up to 100%, but only for NWs with diameters below approx. 7nm. Orientation also plays an important role at these dimensions. Above 7nm diameter, however, both the power factor and *ZT* values saturate, while orientation effects are also smoothened out by some degree. Power factor enhancements are, however, not achievable in all cases. Several subbands very closely packed in energy, help in increasing *S*, whereas light mass subbands enhance *σ*. A combination of both can provide benefits in the power factor, and subband engineering techniques can be employed towards this task.

## Acknowledgements

This work was supported by the European Science Foundation EUROCORES Programme FoNE - DEWINT.

Figure captions

Figure 1:

Zincblende lattice of cylindrical nanowires in [100], [110], and [111] orientations.

Figure 2

Dispersions of n-type NWs in various orientations and diameters ($D$). The left column is for D=3nm wires, and the right column for $D$=12nm wires. (a) [100], $D$=3nm. (b) [100], $D$=12nm. (c) [110], $D$=3nm. (d) [110], $D$=12nm. (e) [111], $D$=3nm. (f) [111], $D$=12nm. $a_0$, $a_0$' and $a_0$'' are the unit cell lengths for the wires in the [100], [110], and [111] orientations, respectively.

Figure 3

Dispersions of p-type NWs in various orientations and diameters ($D$). The left column is for $D$=3nm wires, and the right column for D=12nm wires. (a) [100], $D$=3nm. (b) [100], $D$=12nm. (c) [110], $D$=3nm. (d) [110], $D$=12nm. (e) [111], $D$=3nm. (f) [111], $D$=12nm. $a_0$, $a_0$' and $a_0$'' are the unit cell lengths for the wires in the [100], [110], and [111] orientations, respectively.

Figure 4

Thermoelectric features for the n-type [100] NWs with diameters from $D$=3nm (black-solid line) to $D$=12nm (black-dotted line) as a function of the one-dimensional carrier density (corresponding doping concentration). The arrows indicate the direction of diameter increase. The results are presented with 1nm diameter increment. (a) Electronic conductivity $\sigma$. (b) Seebeck coefficient $S$. (c) Power factor $\sigma S^2$.

Figure 5:

Thermoelectric features for the n-type, and p-type, in [100], [110], and [111] orientations. NWs as a function of the NW diameter. (a) Maximum power factor $\sigma S^2$. (b) Electronic conductivity $\sigma$, at the maximum power factor. (c) Seebeck coefficient $S$, at the maximum power factor.

Figure 6:

$ZT$ figure of merit for the $D$=3nm and $D$=12nm NW devices in the [100], [110], and [111] orientations. (a) n-type. (b) p-type.

Figure 1: Channels in different orientations

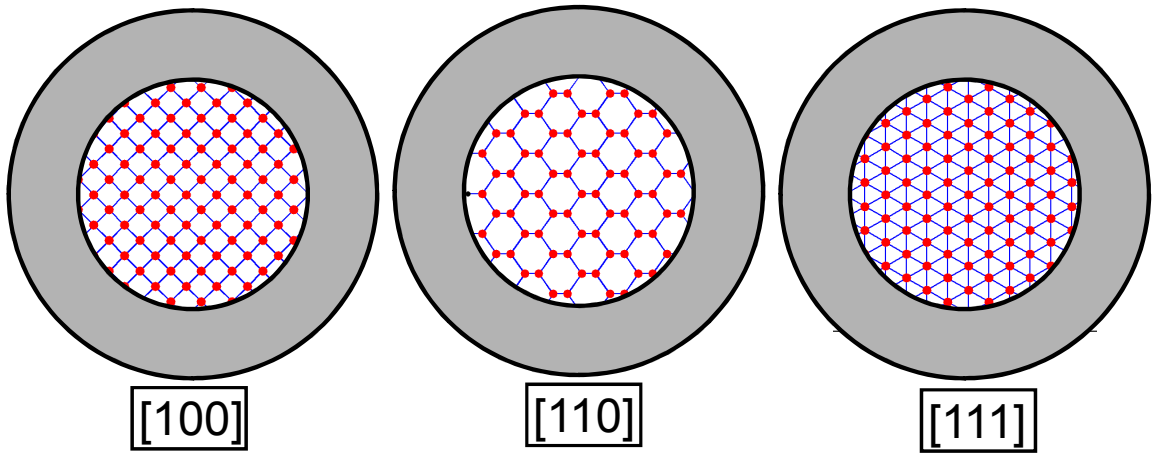

Figure 2: n-type dispersions

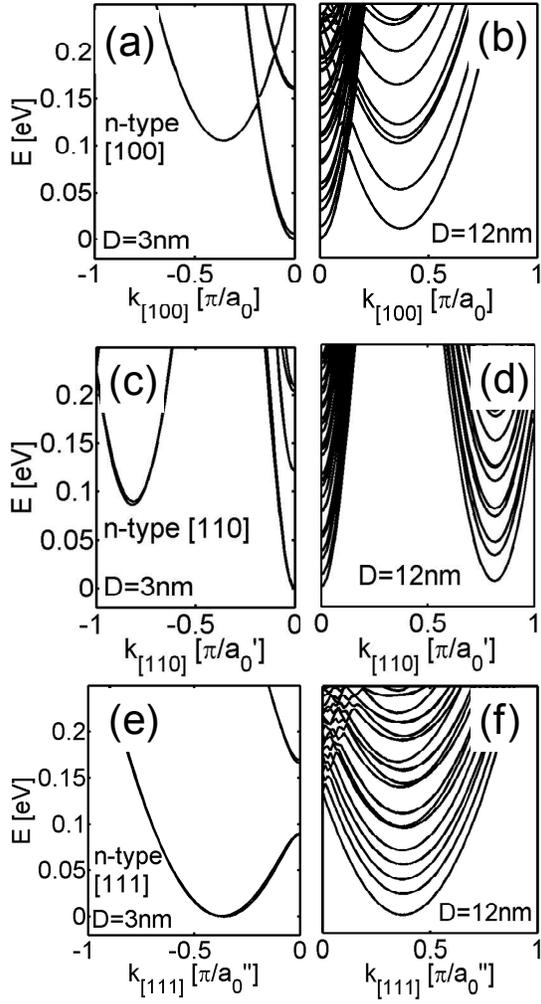

Figure 3: p-type dispersions

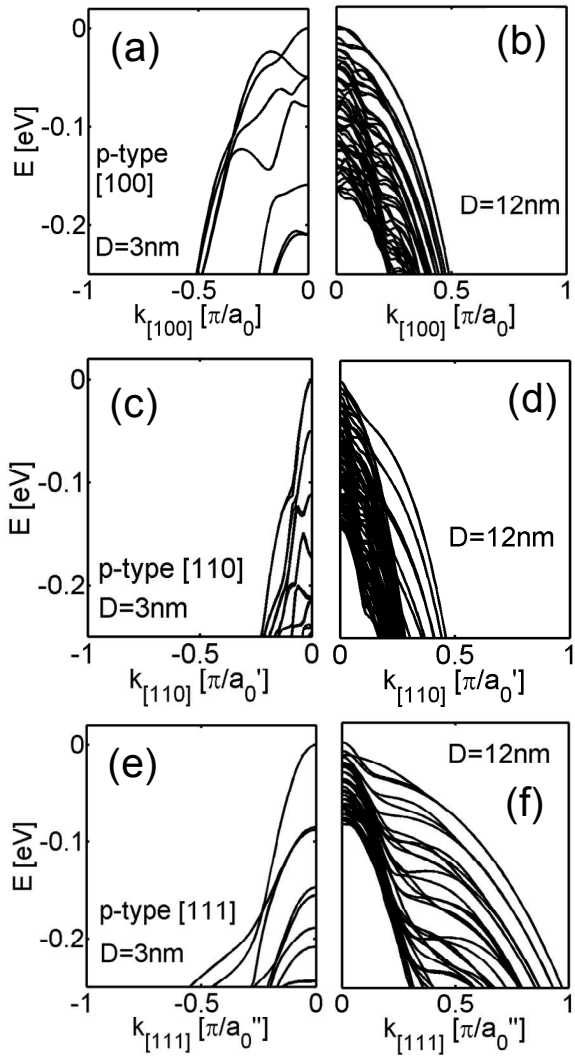

Figure 4: sigma, S, P for n-type [100]

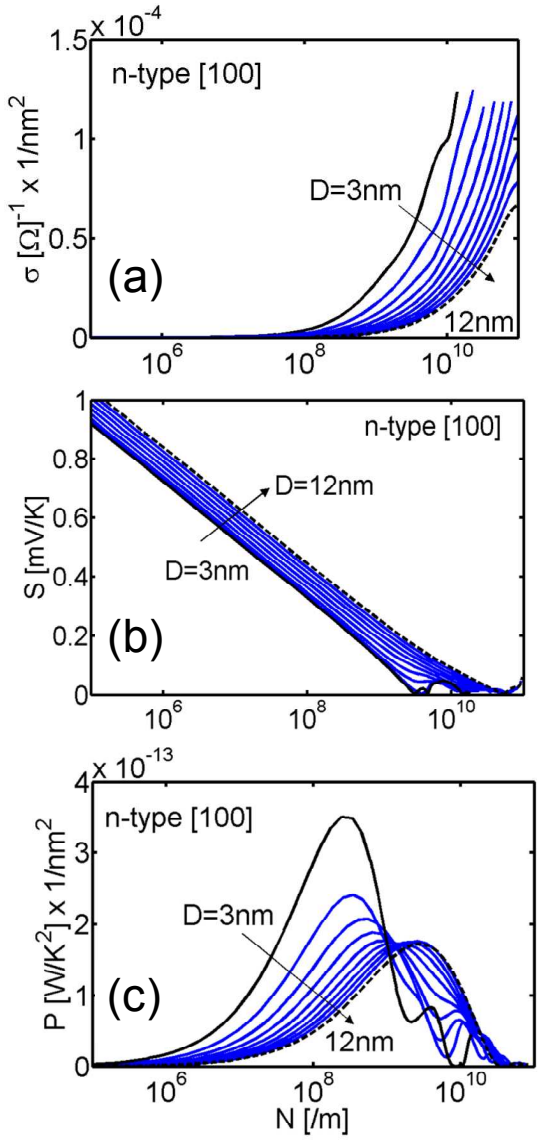

Figure 5: maximum P, sigma and S at max P for all wires

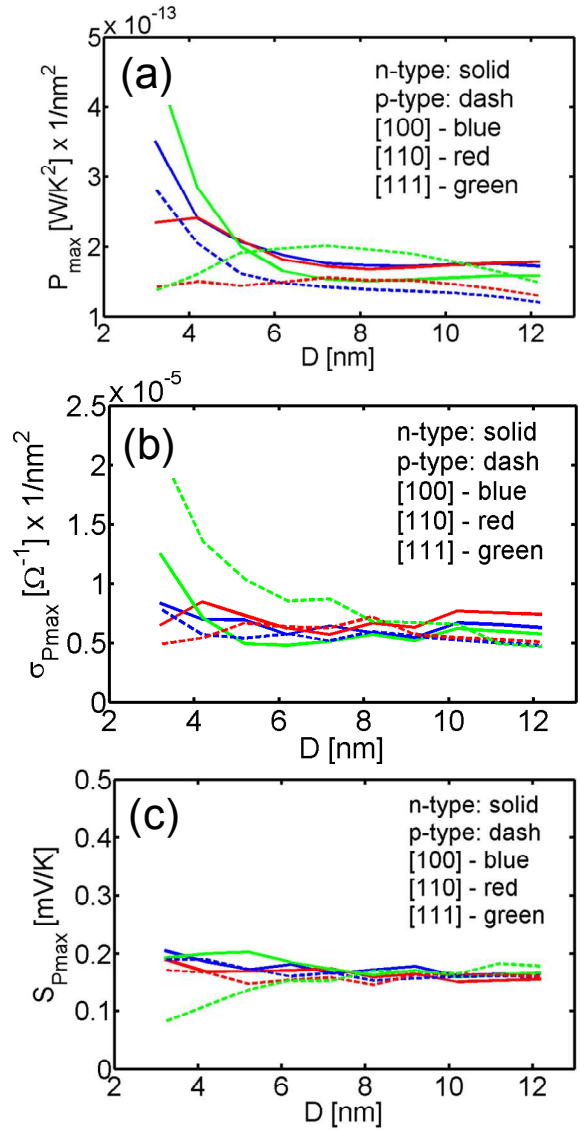

Figure 6: ZT for all devices

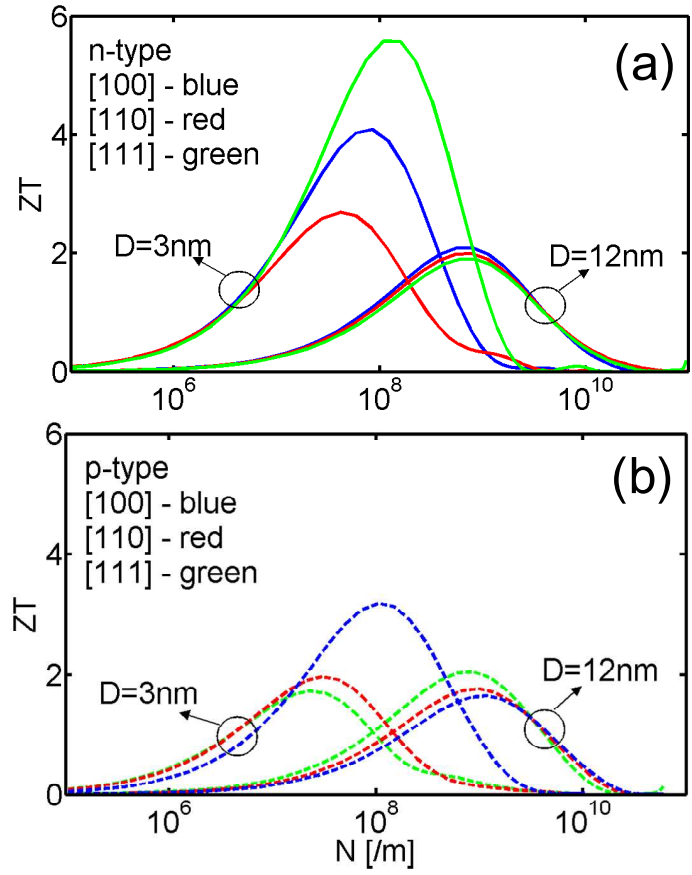